\documentclass[sigconf]{acmart}

\usepackage{color}
\usepackage{bbm}
\usepackage{multirow}
\usepackage[inline]{enumitem}
\usepackage[ruled]{algorithm2e}
\usepackage{booktabs}
\usepackage{caption}
\usepackage{makecell}
\usepackage[skip=0pt]{caption}
\usepackage{amsmath}
\usepackage{hyperref}
\usepackage{array} 
\usepackage{graphicx}
\usepackage{xcolor}
\usepackage{acronym}
\usepackage[export]{adjustbox}
\usepackage{subcaption}
\usepackage{pifont}
\usepackage{siunitx}
\usepackage{colortbl}
\usepackage{balance}
\newcommand{\heading}[1]{\vspace*{0.5mm}\noindent\textbf{#1.}}

\AtBeginDocument{%
  \providecommand\BibTeX{{%
    \normalfont B\kern-0.5em{\scshape i\kern-0.25em b}\kern-0.8em\TeX}}}

\makeatletter
\g@addto@macro\normalsize{%
  \abovedisplayskip 2pt plus1pt 
  \belowdisplayskip 2pt plus1pt
  \abovedisplayshortskip  2pt plus1pt%
  \belowdisplayshortskip  1pt plus1pt
}
\makeatother

\newcommand{\sigp}{\rlap{$^{+}$}}
\newcommand{\sign}{\rlap{$^{-}$}}
\newcommand{\btr}{\rlap{$^{\clubsuit}$}}
\newcommand{\wtr}{\rlap{$^{\diamondsuit}$}}
\newcommand{\btrs}{\rlap{$^{+ \clubsuit}$}}
\newcommand{\wtrs}{\rlap{$^{- \diamondsuit}$}}

\AtBeginDocument{%
  \providecommand\BibTeX{{%
    Bib\TeX}}}

\copyrightyear{2026}
\acmYear{2026}
\setcopyright{cc}
\setcctype{by}
\acmConference[KDD '26]{Proceedings of the 32nd ACM SIGKDD Conference on Knowledge Discovery and Data Mining V.2}{August 09--13, 2026}{Jeju Island, Republic of Korea}
\acmBooktitle{Proceedings of the 32nd ACM SIGKDD Conference on Knowledge Discovery and Data Mining V.2 (KDD '26), August 09--13, 2026, Jeju Island, Republic of Korea}
\acmDOI{10.1145/3770855.3818053}
\acmISBN{979-8-4007-2259-2/2026/08}



\author{Benben Wang}
\authornote{Contributed equally}

\orcid{0009-0003-0630-5710}
\affiliation{
    \institution{State Key Laboratory of AI Safety}
    \institution{Institute of Computing Technology,}
    \institution{Chinese Academy of Sciences}
	\institution{University of Chinese Academy of Sciences}
	\city{Beijing}
	\country{China}
}
\email{benbenwang177@outlook.com}

\author{Minghao Tang}
\authornotemark[1]
\orcid{0009-0002-1911-5142}
\affiliation{
    \institution{State Key Laboratory of AI Safety}
    \institution{Institute of Computing Technology,}
    \institution{Chinese Academy of Sciences}
	\institution{University of Chinese Academy of Sciences}
	\city{Beijing}
	\country{China}
}
\email{tangminghao25s@ict.ac.cn}

\author{Hengran Zhang}
\orcid{0009-0004-1144-1298}
\affiliation{
    \institution{State Key Laboratory of AI Safety}
    \institution{Institute of Computing Technology,}
    \institution{Chinese Academy of Sciences}
	\institution{University of Chinese Academy of Sciences}
	\city{Beijing}
	\country{China}
}
\email{zhanghengran22z@ict.ac.cn}

\author{Jiafeng Guo}
\orcid{0000-0002-9509-8674}
\affiliation{
    \institution{State Key Laboratory of AI Safety}
    \institution{Institute of Computing Technology,}
    \institution{Chinese Academy of Sciences}
	\institution{University of Chinese Academy of Sciences}
	\city{Beijing}
	\country{China}
}
\email{guojiafeng@ict.ac.cn}

\author{Keping Bi}
\orcid{0000-0001-5123-4999}
\authornote{Corresponding author}
\affiliation{
    \institution{State Key Laboratory of AI Safety}
    \institution{Institute of Computing Technology,}
    \institution{Chinese Academy of Sciences}
	\institution{University of Chinese Academy of Sciences}
	\city{Beijing}
	\country{China}
}
\email{bikeping@ict.ac.cn}

\renewcommand{\shortauthors}{Benben Wang, Minghao Tang, Hengran Zhang, Jiafeng Guo, \& Keping Bi}
\begin{document}
\title{Training Dense Retrievers with Multiple Positive Passages}




\begin{abstract}

Modern knowledge-intensive systems, such as retrieval-augmented generation (RAG), rely on effective retrievers to establish the performance ceiling for downstream modules. However, retriever training has been bottlenecked by sparse, single-positive annotations, which lead to false-negative noise and suboptimal supervision. While the advent of large language models (LLMs) makes it feasible to collect comprehensive multi-positive relevance labels at scale, the optimal strategy for incorporating these dense signals into training remains poorly understood. 
In this paper, we present a systematic study of multi-positive optimization objectives for retriever training. We unify representative objectives, including Joint Likelihood (JointLH), Summed Marginal Likelihood (SumMargLH), and Log-Sum-Exp Pairwise (LSEPair) loss, under a shared contrastive learning framework. Our theoretical analysis characterizes their distinct gradient behaviors, revealing how each allocates probability mass across positive document sets. Empirically, we conduct extensive evaluations on Natural Questions, MS~MARCO, and the BEIR benchmark across two realistic regimes: homogeneous LLM-annotated data and heterogeneous mixtures of human and LLM labels. Our results show that LSEPair consistently achieves superior robustness and performance across settings, while JointLH and SumMargLH exhibit high sensitivity to the quality of positives. Furthermore, we find that the simple strategy of random sampling (Rand1LH) serves as a reliable baseline. By aligning theoretical insights with empirical findings, we provide practical design principles for leveraging dense, LLM-augmented supervision to enhance retriever effectiveness.
\end{abstract}

\maketitle

\newcommand\kddcodeurl{https://doi.org/10.5281/zenodo.20323735}
\newcommand\kdddataurl{https://doi.org/10.57967/hf/8883}

\ifboolexpr{not test {\ifdefempty{\kddcodeurl}} or not test {\ifdefempty{\kdddataurl}}}{

\begingroup\small\noindent\raggedright\textbf{Resource Availability:}
\ifdefempty{\kddcodeurl}{}{The source code of this paper is available at \url{\kddcodeurl} and \url{https://github.com/Trustworthy-Information-Access/Multi-Positive-Passages}. }
\ifdefempty{\kdddataurl}{}{The datasets used in this paper are available at \url{\kdddataurl}.}

\endgroup

}

\section{Introduction}

Retrieval aims to identify as many relevant documents as possible from a large corpus and rank them at the top.
It serves as a fundamental component in many knowledge-intensive systems, including web search pipelines with re-ranking stages~\cite{liu2017cascade, glass2022re2g,wang2012extracting} and retrieval-augmented generation (RAG)~\cite{lewis2020retrieval,izacard2023atlas,ram2023context}.
Because downstream modules—such as rerankers or generators—can only operate on retrieved candidates, retriever effectiveness directly determines the upper bound of end-task performance~\cite{shi2024replug,tang2025injecting,zamani2022retrieval}. 
Consequently, improving retrieval quality remains a central problem in information retrieval.

Training effective retrievers, however, critically depends on the availability of relevance supervision.
Obtaining high-quality relevance annotations over large corpora is notoriously expensive: early ad-hoc retrieval benchmarks collected judgments for only a few dozen queries per year~\cite{voorhees2003overview,craswell2020overviewtrec2019deep,craswell2021overviewtrec2020deep}, while large-scale datasets such as MS~MARCO~\cite{nguyen2016ms} typically provide a single positive document per query. 
Such sparse annotation is inherently incomplete. 
It inevitably leads to a scenario where un-annotated relevant documents are treated as false negatives, which confuses the retriever during training~\cite{zhang2025utility,qu2021rocketqa,cai2022hard} and limits the signals available to guide the model.
\looseness=-1

Recent advances in large language models (LLMs) significantly change this landscape. 
LLMs have demonstrated strong capability in performing relevance assessment at scale~\cite{qin2024large,zhang2025utility,yu2025can}, making it feasible to collect richer and more comprehensive relevance labels for retriever training. 
For a long time, the standard training objective has been the optimization of the softmax likelihood of a single positive given a query using contrastive noise estimation including hard negatives and in-batch negatives (SingleLH), or often referred to as InfoNCE~\cite{oord2018representation}. 
However, this objective is not natively suited for multi-positive training. 
It is important to note that while increasing the number of positives in the training batch helps avoid confusion from false negatives, it also decreases the number of hard negatives and causes the candidate distribution to diverge from real-world scenarios where only a small number of positives exist amidst millions of irrelevant documents. 
This divergence can potentially harm model performance, raising the critical question: how should multiple positive documents be effectively incorporated into retriever training?

There are several ways to incorporate multiple positives into the retriever training process.
A straightforward approach is to randomly sample one positive at each iteration and apply the standard InfoNCE objective (Rand1LH). 
Other alternatives include optimizing the joint likelihood of multiple positives occurring together (JointLH) or optimizing the summed marginal likelihood of the positives (SumMargLH). 
Among these listwise methods, JointLH is notably more sensitive to the number and quality of positives because it forces the model to optimize towards all targets simultaneously. 
In contrast, Rand1LH and SumMargLH provide more flexibility~\cite{zhang2025utility}.
Beyond these, pairwise loss functions offer another alternative by optimizing all positive and negative pairs independently~\cite{faysse2024colpali,huang2025beyond}, making them less sensitive to the count of positives. 
The optimization of log-sum-exp over pairwise score differences (LSEPair) is a concrete instance.
Despite their feasibility, these objectives are typically studied in isolation, and their theoretical relationships, empirical behavior, and practical trade-offs remain insufficiently understood.

To fill the gap, in this paper, we present a systematic study of multi-positive optimization objectives for retriever training. 
Theoretically, we unify representative listwise and pairwise objectives under a common contrastive learning framework and analyze their relationships and gradient behaviors. 
We show that JointLH introduces an implicit regularization that encourages uniform allocation of probability mass across positives, SumMargLH concentrates on a small number of dominant positives, and the pairwise LSEPair enforces strict separation between every positive–negative pair, yielding a stronger optimization criterion. 
When there is only a single positive, all the objectives regress to the commonly used InfoNCE.
\looseness=-1

Empirically, we provide comparisons under two representative settings: one where only LLM annotations of similar quality are available, and another where high-quality human annotations are mixed with lower-quality LLM labels. 
We conduct extensive experiments on Natural Questions~\cite{kwiatkowski2019natural}, MS MARCO~\cite{nguyen2016ms}, and out-of-domain BEIR~\cite{thakur2021beirheterogenousbenchmarkzeroshot} benchmarks. 
We find that LSEPair performs robustly and achieves the best performance across most settings. 
In contrast, JointLH and SumMargLH behave unstably when the positive quality is heterogeneous or homogeneous. 
Rand1LH, despite its simplicity, emerges as a strong and reliable alternative under different settings. Importantly, these empirical trends closely align with our theoretical analysis, demonstrating that differences in performance can be explained by how each objective allocates the gradient signal across the positive set.

As LLM-based relevance annotation becomes increasingly feasible, an important practical question is how to allocate a limited annotation budget. In particular, it is unclear whether labeling more queries with fewer positives or fewer queries with richer per-query supervision is more effective for training retrievers. Using the total number of positive labels as a proxy for annotation cost, our results suggest that broader query coverage is more beneficial when annotation is expensive, whereas low-cost LLM annotation enables gains from increasing both query coverage and per-query annotation depth.

In summary, this work makes three primary contributions:
\begin{itemize}[leftmargin=*,itemsep=0pt,topsep=0pt,parsep=0pt]
    \item We unify representative multi-positive optimization objectives within a shared contrastive learning framework, revealing their mathematical relationships and distinct gradient behaviors over positive sets.
    \item We conduct systematic experiments on Natural Questions, MS MARCO, and BEIR under both homogeneous LLM-labeled and heterogeneous human–LLM supervision regimes.
    \item We derive practical guidance by showing that LSEPair is the most robust objective and that empirical stability directly follows from its gradient properties, informing effective use of LLM-augmented supervision.
\end{itemize}

\section{Related Work}
\label{sec:related_work}

\subsection{Dense Retrieval}
\label{sec:related_dr}

Dense retrieval has emerged as a dominant paradigm in modern information retrieval, typically employing dual-encoder architectures initialized from pre-trained language models (PLMs)~\cite{devlin2019bert, karpukhin2020dense, xiao2022retromae}. 
Since PLMs are originally optimized for token-level masked language modeling, considerable effort has been devoted to adapting them for retrieval tasks.
One line of work focuses on retrieval-oriented pre-training~\cite{izacard2021unsupervised,liu2023retromae,wang2023simlm,gao2021condenser}. 
Early approaches bridged this gap by constructing pseudo-relevance signals—for instance, through inverse cloze tasks~\cite{lee2019latent} or contrastive span prediction~\cite{izacard2021unsupervised}—to align query and document embedding spaces.
Other works, such as RetroMAE~\cite{xiao2022retromae} and SimLM~\cite{wang2023simlm}, utilize auto-encoder architectures with shallow decoders to enforce robust sentence embedding learning, significantly improving zero-shot performance.

Regarding the fine-tuning stage, advancements have been largely driven by optimizing the contrastive learning framework.
A critical component is negative sampling, which introduces challenging ``hard'' negatives that closely resemble relevant documents to sharpen the model's discriminative capacity~\cite{cai2022hard,xiong2020approximate,qu2021rocketqa,zhan2021optimizing}. ANCE~\cite{xiong2020approximate} introduced global hard negative mining via an asynchronously updated index, while RocketQA~\cite{qu2021rocketqa} further advanced the paradigm through cross-batch negatives and denoised hard negative sampling, improving both training efficiency and sample quality.
Additionally, knowledge distillation is widely adopted, wherein powerful yet expensive cross-encoders serve as teachers to guide the training of efficient dual-encoders~\cite{qu2021rocketqa,liu2023retromae}.
In this work, we focus on the fine-tuning stage within the contrastive learning framework. 
Distinct from the predominant focus on optimizing negative sampling strategies, we shift our attention to the utilization of positive instances.
Specifically, we investigate how multiple positive documents can be effectively incorporated into the contrastive objective, examining whether these enriched signals can strengthen the learning process and improve retrieval performance.
%

\subsection{Ranking Optimization Objective}
\label{sec:related_ranking_obj}

The ranking optimization objective is critical for learning effective retrieval models, as it dictates how the model distinguishes relevant documents from irrelevant ones in the embedding space.
Historically, optimization objectives are categorized into point-wise, pairwise, and listwise formulations.
(1) Point-wise objectives formulate ranking as a regression or classification problem. They calculate the loss for each query-document pair independently, aiming to minimize the discrepancy between the predicted relevance score and the ground truth label~\cite{nogueira2019passage,li2007mcrank}.
(2) Pairwise objectives, exemplified by RankNet~\cite{burges2005learning}, optimize the relative ordering of document pairs. These methods focus on minimizing the number of inversions, specifically penalizing instances where a negative document outscores a positive one. 
This formulation naturally accommodates multiple positives: given several relevant documents for a query, margin- or hinge-style pairwise losses can be applied to all positive-negative pairs without requiring a single positive target.
To enhance optimization stability and differentiability, smooth surrogate loss functions, such as the Log-Sum-Exp Pairwise (LSEPair) loss~\cite{li2017improving}, have been introduced to approximate non-smooth ranking metrics.
(3) Listwise objectives, such as ListNet~\cite{cao2007learning} and ListMLE~\cite{xia2008listwise}, operate on the entire candidate list collectively. 
Instead of focusing on local pairs, these approaches optimize the probability distribution of permutations or top-1 candidates, aiming to align the predicted ranking list globally with the ground truth.

In dense retrieval, the contrastive InfoNCE loss~\cite{oord2018representation} has become the prevailing optimization objective~\cite{karpukhin2020dense,xiao2022retromae,gao2021condenser,qu2021rocketqa}. 
Theoretically, InfoNCE operates as a specific instance of a listwise objective: given a query and a candidate list containing one positive document alongside multiple negatives, it optimizes the model to maximize the softmax-normalized score assigned to the positive instance.
The standard InfoNCE formulation is constrained to a single-positive setup~\cite{zhang2025utility}. 
In this work, we extend this contrastive framework to accommodate multi-positive scenarios. 
Although multi-positive supervision has been widely considered in general ranking losses, dense retrieval differs in that training is performed over a sampled candidate pool with hard and in-batch negatives, whereas test-time retrieval ranks a query against a much larger corpus with an extremely sparse positive distribution. 
This train-test distribution gap makes the positive-negative ratio and the way positives are aggregated especially important.
By incorporating multiple relevant documents into the objective function, we aim to fully exploit the abundant positive signals, thereby enhancing representation learning and retrieval effectiveness.
%
\section{Multi-Positive Dense Retriever Training}
\label{sec:method}

In this section, we first formalize the dense retrieval task and review the commonly used single-positive contrastive training objective. We then introduce several loss functions for multi-positive retriever training, covering both listwise and pairwise formulations. Finally, we analyze the relationships among these objectives and discuss their theoretical implications.
\subsection{Preliminary}
\label{sec:preliminary}

\heading{Dense Retrieval}
Dense retrieval typically employs a dual-encoder architecture to map queries and documents into a shared latent space.
Let $E_q(\cdot)$ and $E_d(\cdot)$ denote the query and document encoders, respectively.
Given a query $q$ and a document $d$, their fixed-length dense representations are encoded as $\mathbf{q} = E_q(q)$ and $\mathbf{d} = E_d(d)$.
The relevance score $s(q, d)$ is computed via a similarity function $\phi(\cdot, \cdot)$:
\begin{equation}
s(q, d) = \phi(\mathbf{q}, \mathbf{d}).
\end{equation}
In practice, $\phi$ is commonly implemented as the dot product.
The encoders are typically initialized from PLMs, with parameters often shared between $E_q$ and $E_d$. 

\heading{Training Objective}
During training, dense retrievers utilize contrastive learning to distinguish relevant documents from irrelevant ones.
For a query $q$, a candidate pool $\mathcal{D}$ is constructed, comprising a set of positive documents $D^+$ and a set of negative documents $D^-$. $D^-$ typically includes both hard negatives (mined specifically for $q$) and in-batch negatives (from other queries in the batch).
In the common single-positive setting, $D^+$ contains a unique positive $d^+$ (i.e., $\mathcal{D} = \{d^+\} \cup D^-$).
The probability of any document $d \in \mathcal{D}$ to be positive is calculated as:
\begin{equation}
P(d|q, \mathcal{D}) = \frac{\exp(s(q, d))}{\sum_{d' \in \mathcal{D}} \exp(s(q, d'))}.
\end{equation}
The training objective is to maximize the likelihood of the annotated positive $d^+$. 
We refer to this single-positive likelihood objective as \textbf{SingleLH}, whose loss is the negative log-likelihood:
\begin{equation}
\mathcal{L}_{\text{SingleLH}}(q, d^+, D^-) = - \log P(d^+|q, \mathcal{D}).
\end{equation}
This formulation is equivalent to maximizing the mutual information between the query and the positive document, with the softmax denominator approximated via contrastive noise estimation—commonly referred to as the InfoNCE loss~\cite{oord2018representation}.
\begin{equation}
\label{eq:infonce}
\mathcal{L}_{\text{InfoNCE}} = - \log \frac{\exp(s(q, d^+))}{\exp(s(q, d^+)) + \sum_{d^- \in D^-} \exp(s(q, d^-))}.
\end{equation}
For clarity, we summarize the critical notations used throughout this paper in Table~\ref{tab:notations}.

\begin{table}[t]
\centering
\caption{Summary of critical notations.}
\renewcommand{\arraystretch}{0.9}
\label{tab:notations}
\resizebox{\columnwidth}{!}{%
\begin{tabular}{l|l}
    \toprule
    \textbf{Notation} & \textbf{Description} \\
    \midrule
    $q, d$ & A query and a document instance \\
    $\mathbf{q}, \mathbf{d}$ & Dense embeddings of query $q$ and document $d$ \\
    $s(q, d)$ & Relevance score (similarity) between $q$ and $d$ \\
    \midrule
    
    $D^+$ & Set of positive documents for query $q$ \\
    $D^-$ & Set of negative documents for query $q$ \\
    $\mathcal{D}$ & Candidate document pool for training, $\mathcal{D} = D^+ \cup D^-$ \\
    \midrule
    
    $P(d|q, \mathcal{D})$ & Probability of selecting document $d$ from $\mathcal{D}$ \\
    $\mathcal{L}$ & Training objective (Loss function) \\
    \bottomrule
\end{tabular}%
}
\end{table}

\subsection{Multi-Positive Objectives}

To effectively incorporate multiple positive documents into retriever training, we extend the single-positive objective SingleLH (i.e., InfoNCE) to the multi-positive setting. 
Building upon this foundation, we derive four distinct loss variants tailored to accommodate an expanded positive set $D^+$ with $|D^+| > 1$.
These variants represent different strategies for aggregating relevance signals within the contrastive learning framework.

\heading{Rand1LH}
The most straightforward adaptation is to randomly sample a single positive document $d^+_i$ from the set $D^+$ during each epoch and optimize its likelihood.
We refer to this method as \textit{Rand1LH}.
By treating the sampled $d^+_i$ as the sole ground truth for that step, it directly employs the standard InfoNCE formulation:
\begin{equation}
\mathcal{L}_{\text{Rand1LH}} = - \log \frac{\exp(s(q, d^+_i))}{\exp(s(q, d^+_i)) + \sum_{d^- \in D^-} \exp(s(q, d^-))},
\end{equation}
where $d^+_i$ is drawn from a uniform distribution $\mathcal{U}(D^+)$.

This approach requires minimal modification to the existing training pipeline or hyperparameters, as it maintains the same positive-negative ratio in the training batch as the typical single-positive setting.
However, processing positive signals in isolation rather than jointly may fail to fully exploit the enriched supervision, particularly when training epochs are limited.

\heading{JointLH}
Another natural adaptation is to optimize the joint likelihood of all available positive passages within a single iteration.
We denote this approach as \textit{JointLH}.
Assuming the relevance of each positive document is conditionally independent given the query, the objective can be formulated as the average negative log-likelihood across the positive set $D^+$:
\begin{equation}
\begin{aligned}
\mathcal{L}_{\text{JointLH}} &= - \frac{1}{|D^+|} \sum_{d^+ \in D^+} \log P(d^+|q, \mathcal{D}) \\
&= - \frac{1}{|D^+|} \sum_{d^+ \in D^+} \log \frac{\exp(s(q, d^+))}{\sum_{d' \in \mathcal{D}} \exp(s(q, d'))}.
\end{aligned}
\end{equation}
Here, $|D^+|$ denotes the number of positive passages.
This objective leverages the full set of relevance signals simultaneously, encouraging the model to assign high scores to every ground-truth passage relative to negatives in $\mathcal{D}$.
This enforces a strict constraint requiring all positives to achieve high relevance scores, which can be problematic when the positive set contains noise, such as false positives or marginally relevant passages.

\heading{SumMargLH}
In contrast to optimizing the joint likelihood of the positive set, this approach maximizes the summed marginal likelihood of the documents in the set \cite{zhang2025utility}. 
We refer to this approach as \textit{SumMargLH}.
Instead of optimizing all individual probabilities, it maximizes the cumulative probability mass of the entire set $D^+$:
\begin{equation}
\begin{aligned}
\mathcal{L}_{\text{SumMargLH}} &= - \log \sum_{d^+ \in D^+} P(d^+|q, \mathcal{D}) \\
&= - \log \frac{\sum_{d^+ \in D^+} \exp(s(q, d^+))}{\sum_{d' \in \mathcal{D}} \exp(s(q, d'))}.
\end{aligned}
\end{equation}
This formulation relaxes the optimization objective: it does not require the likelihood of every positive instance to be maximized individually.
Instead, it encourages the model to assign high aggregate probability to the set $D^+$, allowing it to prioritize the most confident positives while being tolerant of potential label noise.

\heading{LSEPair}
Alternatively, we can shift our perspective from maximizing softmax probabilities to optimizing the relative ordering between positive and negative pairs.
First, we observe that the standard SingleLH (i.e., InfoNCE) in Equation \eqref{eq:infonce} can be mathematically rewritten as a function of score differences:
\begin{equation}
\label{eq:infonce_pair}
\mathcal{L}_{\text{SingleLH}} = \log \left( 1 + \sum_{d^- \in D^-} \exp(s(q, d^-) - s(q, d^+)) \right).
\end{equation}
This formulation reveals that InfoNCE essentially aggregates the pairwise score differences between the single positive and all negatives in the candidate set $\mathcal{D}$.
Motivated by this, we extend the aggregation scope to encompass all pairs of positive and negative documents.
This leads to the Log-Sum-Exp Pairwise loss (\textit{LSEPair}), originally proposed for multi-label classification~\cite{li2017improving}, which we adapt for dense retrieval:
\begin{equation}
\label{eq:lsep}
\mathcal{L}_{\text{LSEPair}} = \log \left( 1 + \sum_{d^+ \in D^+} \sum_{d^- \in D^-} \exp(s(q, d^-) - s(q, d^+)) \right).
\end{equation}
By summing over the Cartesian product of $D^+$ and $D^-$, LSEPair explicitly penalizes any case where a negative document scores higher than a positive one, enforcing a robust separation between the relevant and irrelevant sets.

\heading{Dynamic Positive Reweighting}
Beyond these fixed aggregation strategies, the contribution of each positive can also be dynamically weighted according to its estimated relevance to the query. For example, ReContriever~\cite{lei-etal-2023-unsupervised} estimates the reliability of pseudo-positive pairs during unsupervised dense retrieval pre-training and uses the estimated relevance to reweight the contrastive loss. This setting differs from ours, because we study supervised dense retrieval with explicitly labeled multi-positive annotations. Nevertheless, to provide a more complete comparison, we adapt this idea as a dynamic positive reweighting baseline and report the results in Appendix~\ref{sec:appendix_dynamic_weighting}.
\subsection{Objective Characteristic and Connection}
\label{sec:methods:properties}
\heading{Regression to SingleLH} 
All four multi-positive objectives are natural extensions of the SingleLH formulation.
When the positive set contains a unique document (i.e., $|D^+| = 1$), all four variants mathematically regress to the standard SingleLH (InfoNCE) loss.

\heading{JointLH Equalizes Positives' Probability}
Gradient analysis reveals that JointLH implicitly enforces uniform probability allocation across all positive documents.
Specifically, the gradient with respect to the score of a positive instance $d^+_i \in D^+$ is given by:
\begin{equation}
\frac{\partial \mathcal{L}_{\text{JointLH}}}{\partial s(q, d^+_i)} = P(d^+_i|q, \mathcal{D}) - \frac{1}{|D^+|}.
\label{eq:joint_property}
\end{equation}
This formulation drives the optimization toward an equilibrium in which each positive receives probability mass $1/|D^+|$.
Consequently, when positive quality varies within a batch, and a high-quality positive attains a probability $(P(d_i^+) > 1/|D^+|)$, JointLH will suppress its score through gradient descent. Conversely, lower-quality positives with probabilities below $1/|D^+|$ will be pushed upward toward this uniform target. This behavior can be problematic when positive labels exhibit heterogeneous quality, as it may over-promote weaker positives and dampen strong ones. However, when positive quality is relatively homogeneous, this implicit balancing effect can strengthen the optimization of all positives, encouraging broader coverage of relevant documents and potentially improving recall.

\heading{SumMargLH Emphasizes High-scoring Positives}
Unlike JointLH, SumMargLH allocates supervision signals non-uniformly via an implicit re-weighting mechanism.
Specifically, the gradient with respect to a positive instance $d^+_i \in D^+$ can be factorized into a shared global error term and a positive-specific local weight:
\begin{equation}
\label{eq:summar-grad}
\small
\frac{\partial \mathcal{L}_{\text{SumMargLH}}}{\partial s(q, d^+_i)} \!\!=\!\!
\underbrace{\frac{\exp(s(q, d^+_i))}{\sum_{d' \in \mathcal{D}} \exp(s(q, d'))}}_{\text{Local Weight}} \!\cdot \underbrace{\!\left( 1 \!\! - \!\! \frac{\sum_{d' \in \mathcal{D}} \exp(s(q, d'))}{\sum_{d' \in D^+} \exp(s(q, d'))}\!\right)}_{\text{Global Error Term}}.
\end{equation}
This formulation reveals that the gradient for each positive is directly proportional to its local weight.
Consequently, the highest-scoring positive dominates the gradient update, suppressing the influence of low-scoring or noisy positives.
However, it may also lead to the under-utilization of supervision, as the model tends to focus on the ``easiest'' positive, neglecting other valid signals.

\heading{LSEPair Emphasizes Low-Scoring Positives}
Analysis of the LSEPair gradient reveals a distinct optimization dynamic: it prioritizes ``hard'' positives (i.e., those with lower relevance scores).
Let $Z = 1 + \sum_{d^+_j \in D^+} \sum_{d^- \in D^-} \exp(s(q, d^-) - s(q, d^+_j))$ be the normalization term.
The gradient with respect to a positive $d^+_i$ can be factorized as:
\begin{equation}
\label{eq:lse_grad}
\frac{\partial \mathcal{L}_{\text{LSEPair}}}{\partial s(q, d^+_i)} = - \underbrace{\exp(-s(q, d^+_i))}_{\text{Local Weight}} \cdot \underbrace{\frac{\sum_{d^- \in D^-} \exp(s(q, d^-))}{Z}}_{\text{Global Error Term}}.
\end{equation}
This shows that the gradient scales with $\exp(-s(q, d^+_i))$, meaning lower-scoring positives receive stronger gradients.
Unlike SumMargLH, this mechanism forces the model to attend to the most challenging relevant passages, enforcing a strict separation between every positive–negative pair.
Nevertheless, the same hard-positive emphasis can make LSEPair more sensitive to noisy positives: if a false positive receives a low score, the objective will assign it a large gradient and push it away from negatives as if it were relevant.

\heading{Rand1LH is a Stochastic Approximation of LSEPair}
Comparing Equations~\eqref{eq:infonce_pair} and~\eqref{eq:lsep} reveals that LSEPair explicitly aggregates the pairwise constraints that Rand1LH samples individually.
Consequently, Rand1LH stochastically approximates LSEPair over sufficient training iterations.
However, they diverge in per-step dynamics: LSEPair aggregates all positive signals to reduce gradient variance, whereas Rand1LH preserves the standard positive-to-negative ratio at the cost of higher stochasticity.

\heading{Sensitivity to the Positive–Negative Ratio}
Including more positives in a training batch increases the proportion of positives in the candidate set, causing the training distribution to deviate further from realistic retrieval scenarios in which only a small number of positives exist among a vast pool of irrelevant documents. At the same time, the number of hard negatives is reduced. Both effects can adversely impact retriever performance. Among the objectives we study, JointLH should be particularly sensitive to the positive–negative ratio, as its listwise formulations directly depend on the composition of the candidate set and all the positives contribute to the loss equally. Although SumMargLH is also a list-wise loss, it may be less affected by the number of positives, since it emphasizes high-scoring positives and can remain relatively stable when additional positives are weak. In contrast, Rand1LH and LSEPair could be more robust: Rand1LH preserves the same one-positive-versus-$|\mathcal{D}|$ training ratio as standard single-positive InfoNCE, while the pairwise LSEPair objective optimizes positive–negative score differences independently, making it less sensitive to distributional shifts than listwise losses.

\section{Experimental Setup}
\label{sec:exp_setup}

\subsection{Datasets and Evaluation Metrics}
We utilize two widely established datasets, MS~MARCO~\cite{nguyen2016ms} and Natural Questions (NQ)~\cite{kwiatkowski2019natural}, for model training and in-domain evaluation. 
Additionally, we evaluate on the BEIR benchmark~\cite{thakur2021beirheterogenousbenchmarkzeroshot} to assess out-of-domain generalization.

\heading{Natural Questions} NQ~\cite{kwiatkowski2019natural} is a widely adopted benchmark for open-domain question answering. 
We use its original training set containing approximately 58k queries.
To support multi-positive training, we adopt the exhaustive annotations provided by~\citet{zhang2025utility}, as detailed in Section~\ref{sec:annotation}.
For evaluation, we use the standard test set of 3,610 queries and report performance using Top-20 and Top-100 Accuracy.

\heading{MS~MARCO} 
MS~MARCO~\cite{nguyen2016ms} is a large-scale benchmark derived from real-world Bing search logs.
We utilize the standard training set containing approximately 400k queries, which we re-annotate to obtain multiple positive passages, as detailed in Section~\ref{sec:annotation}. 
For evaluation, we report results on the MS~MARCO Dev set (6,980 queries) using MRR@10, Recall@100 and Recall@1000. 
We also evaluate on the TREC DL 2019 (43 queries) and 2020 (54 queries) test sets~\cite{craswell2020overviewtrec2019deep, craswell2021overviewtrec2020deep}, reporting NDCG@10 as the primary metric.

Given that our models are trained on LLM-generated multi-positive annotations, standard benchmarks present limitations: 
the MS~MARCO Dev set suffers from label sparsity, TREC DL test sets are limited in scale, and human–LLM preference misalignment may introduce evaluation bias.
To address these concerns, we additionally adopt the \textbf{Hybrid Annotation} set from \citet{zhang2025utility}, comprising 200 queries sampled from MS~MARCO Dev. 
Its construction pools top-ranked passages from diverse retrievers and employs GPT-4o-mini~\cite{hurst2024gpt} to identify additional positives based on ground-truth answers. 
The final judgments combine human annotations with LLM-verified positives, reducing false negatives and better aligning with our multi-positive training distribution.

\heading{BEIR} 
To assess out-of-domain generalization, we evaluate on the BEIR benchmark~\cite{thakur2021beirheterogenousbenchmarkzeroshot}, a heterogeneous collection spanning multiple retrieval tasks across diverse domains such as biomedicine and finance. 
Following standard practice, we perform zero-shot evaluation on the 14 publicly available datasets using models trained exclusively on MS~MARCO, and report results with NDCG@10 as the primary metric.
%

\subsection{Multi-Positive Annotation Construction}
\label{sec:annotation}

To enable effective multi-positive training, we construct enriched versions of the training datasets where each query is associated with multiple positive passages. 
Specifically, we leverage LLMs to re-annotate the queries. 
The detailed construction pipelines for each dataset are described below.

\heading{NQ} 
We adopt the multi-positive annotations from \citet{zhang2025utility}, constructed via a utility-focused pipeline designed to ensure high-quality relevance labels. 
The pipeline comprises four stages:
(1) Candidate Retrieval: For each query, a candidate pool is formed by merging top-ranked results from multiple unsupervised retrievers (BM25, RetroMAE~\cite{xiao2022retromae}, and LLM-QL~\cite{zhang2025unleashingpowerllmsdense}). 
(2) Relevance Filtering: An LLM (Qwen3-32B~\cite{yang2025qwen3}) performs coarse filtering to retain topically relevant passages from the candidate pool.
(3) Pseudo-Answer Generation: The same LLM generates a pseudo-answer grounded in the filtered passages.
(4) Utility Verification: The LLM selects passages that are useful and necessary for generating the pseudo-answer from relevant passages. 
Passages passing the final verification stage serve as positives, yielding an average of 5.5 positives per query. 
This utility-driven criterion ensures that selected passages not only share topical relevance but also provide information required to answer the query.

\heading{MS~MARCO} 
We apply the same utility-focused annotation pipeline to MS~MARCO as described for NQ. 
For candidate retrieval, we aggregate results from BM25, RetroMAE~\cite{xiao2022retromae}, and Contriever~\cite{izacard2021unsupervised}. 
Subsequent stages—relevance filtering, pseudo-answer generation, and utility verification—are performed using Qwen3-32B~\cite{yang2025qwen3}. 
This process results in an enriched training set with an average of 6.5 positive passages per query.

\heading{Positive Group Construction} 
In real-world retrieval scenarios, the quality of positive signals is often uneven, ranging from consistent synthetic data to heterogeneous mixtures of gold and silver labels.
To investigate how different objectives adapt to such variations, we employ two distinct positive group configurations for training: 
(1) Homogeneous LLM-annotated positive group: groups composed exclusively of LLM-annotated positives with relatively homogeneous quality;
(2) Heterogeneous mixed positive group: groups combining a single high-quality human-annotated positive (placed first) with additional LLM-annotated positives of potentially lower utility.
This design enables a controlled investigation into how objectives handle quality variance within the positive set.

\subsection{Training Setting}
\heading{Training Objectives}
We evaluate different training objectives detailed in Section~\ref{sec:method}, comprising the standard single-positive objective \textbf{SingleLH}, and four multi-positive variants:
(1) \textbf{Rand1LH}, (2) \textbf{JointLH}, (3) \textbf{SumMargLH}, (4) \textbf{LSEPair}. 
For SingleLH, we utilize the first passage from the LLM-annotated positive set for each query. 
Since the annotation pipeline outputs positives in descending order of utility, this choice typically yields a higher-quality positive compared to subsequent candidates.

\heading{Curriculum Learning}
Recent work~\cite{zhang2025utility} shows that when dealing with data of mixed quality (e.g., massive synthetic data vs. scarce human annotations), a curriculum learning (CL) strategy—specifically, training on lower-quality synthetic data first before refining on high-quality labels—significantly outperforms simple data merging.
Motivated by this, we conduct additional controlled experiments to verify the impact of different multi-positive objectives within this framework.
Specifically, we employ a two-stage protocol: the retriever is first trained on the homogeneous LLM-annotated positives using various multi-positive objectives, and subsequently fine-tuned on human-annotated data using standard SingleLH.

\begin{table*}[t]
\centering
\caption{In-domain retrieval performance comparison under homogeneous LLM-annotated positives. \textbf{Bold} and \underline{underline} denote the best and second-best results, respectively. ``+'' and ``-'' indicate statistically significant improvements and drops compared to SingleLH, while $\clubsuit$ and $\diamondsuit$ indicate statistically significant differences compared to Rand1LH (two-sided paired $t$-test, $p < 0.05$).}
\renewcommand{\arraystretch}{0.8}
\label{tab:indomain_results}
\resizebox{\textwidth}{!}{
\begin{tabular}{l cc ccc c c cc}
\toprule
\multirow{3}{*}{\textbf{Objective}} & \multicolumn{2}{c}{\textbf{NQ}} & \multicolumn{6}{c}{\textbf{MS~MARCO}} \\

\cmidrule(lr){2-3} \cmidrule(lr){4-10}
& \multicolumn{2}{c}{\textbf{Test Set}} & \multicolumn{3}{c}{\textbf{Dev Set}} & \textbf{DL19} & \textbf{DL20} & \multicolumn{2}{c}{\textbf{Hybrid Annotation}} \\
\cmidrule(lr){2-3} \cmidrule(lr){4-6} \cmidrule(lr){7-7} \cmidrule(lr){8-8} \cmidrule(lr){9-10}

& \textbf{Acc@20} & \textbf{Acc@100} & \textbf{MRR@10}  & \textbf{Recall@100} & \textbf{Recall@1000} & \textbf{NDCG@10} & \textbf{NDCG@10} & \textbf{MRR@10} & \textbf{NDCG@10} \\
\midrule
SingleLH
& \underline{76.18} & 84.57
& 29.91 & 81.54 & 93.60
& 62.58 & 60.20
& 75.43 & 48.96 \\
\midrule
Rand1LH
& 75.68 & 84.76
& \underline{30.44}\sigp & 82.02 & 94.11\sigp
& \underline{63.09} & 61.15
& \textbf{79.19} & \textbf{53.07}\sigp \\

JointLH
& 75.93 & \underline{85.15}
& 28.22\wtrs & \underline{83.09}\btrs &\textbf{95.10}\btrs
& 59.81 & \textbf{63.20}
& 75.44 & 51.71\sigp \\

SumMargLH
& 75.12\sign & 84.57
& 29.79\wtr & 80.76\wtrs & 92.98\wtrs
& 62.50 & 61.24
& 78.52 & 49.86\wtr \\

LSEPair
& \textbf{77.01}\btr & \textbf{85.62}\rlap{$^{+ \clubsuit}$}
& \textbf{30.57}\sigp & \textbf{83.10}\btrs & \underline{94.82}\btrs
& \textbf{65.33} & \underline{62.80}
& \underline{78.96} & \underline{52.99}\sigp \\
\bottomrule
\end{tabular}
}
\end{table*}

\subsection{Implementation Details}
All models are initialized from \texttt{bert-base-uncased}~\cite{devlin2019bert} and implemented using the Tevatron toolkit~\cite{gao2022tevatron}.
All experiments were conducted on NVIDIA A800 GPUs with 80GB of memory.

\heading{Training Configuration} 
For each query, we set the passage group size to $G = 8$, where each group contains positive passages and hard negatives. 
To balance positive signal utilization and negative discrimination, for models trained with multi-positive objectives, we enforce a constraint where at most $M$ positive passages are included per group (defaulting to $M=4$). 
If fewer than $M$ positives are available, all positives are used, with the remainder of the group filled by hard negatives.
Following standard practice, we also utilize \textbf{in-batch negatives} to further expand the negative pool.

\heading{Hyperparameters}
Training hyperparameters differ slightly across datasets to accommodate their scale.
On NQ, models are trained for 40 epochs with a global batch size of 64 and a learning rate of 1e-5. 
On MS~MARCO, models are trained for 3 epochs with a global batch size of 128 and a learning rate of 3e-5.

For the curriculum learning experiments on MS~MARCO, the first stage (training on LLM data) follows these standard settings.
The second stage (fine-tuning on human data) is conducted for 1 additional epoch, maintaining the identical batch size and learning rate.\looseness=-1

\section{Experimental Results}
\label{sec:results}

\subsection{Performance on Homogeneous Positives}
\label{sec:res_uniform}


\heading{In-Domain Results}
Table~\ref{tab:indomain_results} summarizes in-domain retrieval performance on NQ and MS~MARCO under the homogeneous LLM-annotated positives. 
Under this setting, we observe the following key findings:

\textit{(1) The pairwise LSEPair objective demonstrates the most robust performance across both datasets.} It achieves the best Top20/Top100 scores on NQ, as well as the highest MS MARCO Dev MRR@10 and DL19 NDCG@10, while also improving recall metrics. This aligns with our theoretical derivation: the LSEPair objective enforces strict separation between each positive–negative pair, resulting in a stronger optimization criterion. 

\textit{(2) Rand1LH is a strong and reliable alternative.}
Rand1LH performs on par with the LSEPair objective on several metrics and delivers the best results on the densely annotated Hybrid Annotation set. 
Rand1LH uniformly samples one positive example and optimizes the single-positive InfoNCE loss per training step.
Theoretically, under ideal conditions, Rand1LH shares the same upper bound as LSEPair; however, its optimization is less efficient because it does not jointly optimize multiple positives within a batch as LSEPair does. These experimental results are consistent with the theoretical analysis.

\textit{(3) JointLH and SumMargLH exhibit unstable performance.} 
JointLH prioritizes low-rank retrieval performance over top-rank effectiveness. 
This behavior is consistent with its probability equalization property as shown in Eq.~\eqref{eq:joint_property},
which drives the model toward a uniform allocation of probability mass across positives.
As a result, when the model assigns high confidence to the strongest positive, the gradient acts as a restorative force that suppresses its score. 
This prevents the formation of a peaky ranking distribution, which is essential for maximizing top-ranked results. 
SumMargLH is consistently weaker and sometimes degrades relative to SingleLH.
This aligns with its gradient characteristic: the positive-side gradient is scaled by a within-positive softmax weight, so higher-scoring positives receive disproportionately larger updates.
Consequently, training focuses on the easiest positives, underutilizing supervision signals, and limiting top-ranking performance.  
%

\begin{table}[t]
\centering
\caption{Zero-shot retrieval performance (NDCG@10) on the BEIR benchmark. \textbf{Bold} and \underline{underline} indicate the best and second-best results.}
\label{tab:ood_results}
\renewcommand{\arraystretch}{0.9}
\setlength{\tabcolsep}{1.5pt}
\begin{tabular*}{\columnwidth}{@{\extracolsep{\fill}} l ccccc}
\toprule
\textbf{Datasets} & SingleLH & Rand1LH & JointLH & SumMargLH & LSEPair \\
\midrule
\textbf{ArguAna}   & \underline{34.87} & 28.66 & \textbf{39.52} &  28.40 & 31.79 \\
\textbf{C-FEVER}   & \underline{21.50} & 21.31 & 21.04 & 21.08 & \textbf{22.05} \\
\textbf{CQA}       & \textbf{26.73} & 23.28 & \underline{24.58} & 23.60 & 23.98 \\
\textbf{DBPedia}   & 28.82 & 28.49 & \textbf{30.49} & 27.67 & \underline{29.73} \\
\textbf{FEVER}     & 64.10 & \textbf{64.66} & 61.10 & 63.48 & \underline{64.24} \\
\textbf{FiQA}      & 24.40 & \underline{24.42} & 23.53 & 23.52 & \textbf{24.87} \\
\textbf{HotpotQA}  & \underline{43.02} & 42.20 & 41.19 & 40.68 & \textbf{43.33} \\
\textbf{NFCorpus}  & 24.66 & 24.84 & \underline{25.41} & 23.41 & \textbf{25.92} \\
\textbf{NQ}        & 43.13 & \underline{45.34} & 42.89 & 43.22 & \textbf{45.91} \\
\textbf{Quora}     & \textbf{81.58} & 80.81 & 65.88 & 80.36 & \underline{81.26} \\
\textbf{SCIDOCS}   & \underline{10.32} & 9.83 & 10.27 & 8.62 & \textbf{10.39} \\
\textbf{SciFact}   & 50.35 & 48.64 & \underline{50.60} & 46.27 & \textbf{52.08} \\
\textbf{Touche}    & \underline{65.24} & 63.34 & 62.28 & 59.78 & \textbf{66.14} \\
\textbf{T-COVID}   & 30.72 & \underline{30.80} & 28.65 & 29.33 & \textbf{31.10} \\
\midrule
\textbf{Average} & \underline{39.25} & 38.33 & 37.67 & 37.10 & \textbf{39.49} \\
\bottomrule
\end{tabular*}

\end{table}
\heading{Out-of-Domain Results}
To assess out-of-domain robustness, we evaluate MS~MARCO-trained models on 14 datasets from the BEIR benchmark as shown in Table~\ref{tab:ood_results}.
Notably, LSEPair demonstrates superior generalization, achieving the highest performance on 9 out of 14 datasets and outperforming other training objectives.
Together with the in-domain results, this suggests that explicitly enforcing pairwise ranking constraints leads to more accurate relevance discrimination, yielding stable gains in both in-domain and out-of-domain settings.
In contrast, Rand1LH and listwise aggregation strategies (JointLH and SumMargLH) tend to overfit the source domain's specific relevance distribution, leading to degraded zero-shot transfer performance.

%


\subsection{Performance on Heterogeneous Positives}
\label{sec:res_nonuniform}
%

\begin{table}[t]
\centering
\caption{Performance on MS~MARCO on human and LLM annotated heterogeneous positives. R@1k and N@10 denote Recall@1000 and NDCG@10, respectively. \textbf{Bold} and \underline{underline} indicate the best and second-best results. $\clubsuit$ and $\diamondsuit$ indicate statistically significant differences compared to Rand1LH (two-sided paired $t$-test, $p < 0.05$).}
\renewcommand{\arraystretch}{0.9}
\label{tab:nonuniform_results}
\resizebox{\linewidth}{!}{%
\begin{tabular}{lcccccc}
\toprule
\multirow{2}{*}{\textbf{Objective}} & \multicolumn{2}{c}{\textbf{Dev Set}} & \textbf{DL19} & \textbf{DL20} & \multicolumn{2}{c}{\textbf{Hybrid Annotation}} \\
\cmidrule(lr){2-3} \cmidrule(lr){4-4} \cmidrule(lr){5-5} \cmidrule(lr){6-7}
& \textbf{MRR@10} & \textbf{R@1k} & \textbf{N@10} & \textbf{N@10} & \textbf{MRR@10} & \textbf{N@10} \\
\midrule
Rand1LH & \underline{30.64} & 94.55 & \underline{62.39} & 63.31 & 77.30 & 51.06 \\
JointLH  & 28.45\wtr & \textbf{95.65}\btr & 59.68 & \underline{64.62} & \textbf{78.45} & \underline{52.46} \\
SumMargLH & 30.43 & 93.10\wtr & 60.53 & 60.01\wtr & 74.29 & 46.61\wtr \\
LSEPair   & \textbf{30.68} & \underline{95.52}\btr & \textbf{64.13} & \textbf{65.04} & \underline{78.32} & \textbf{53.14}\btr \\
\bottomrule
\end{tabular}%
}
\end{table}
Table~\ref{tab:nonuniform_results} reveals retrieval performance under heterogeneous positive quality: 
\textit{(1) LSEPair achieves the best or second-best performance across all metrics,} which further indicates that the LSEPair performs robustly. 
\textit{(2) SumMargLH excels in top-rank retrieval performance but limits low-rank retrieval performance.} 
When a high-quality human positive is usually used as the primary supervision signal, SumMargLH achieves a competitive MRR@10 of 30.43\% on the Dev set, a notable improvement over its performance in the homogeneous regime, indicating that SumMargLH is more sensitive to the positive quality. 
Similar to the homogeneous positives, it yields the lowest Recall@1000, suggesting that it tends to concentrate the learning on the easiest positive. 
\textit{(3) JointLH prioritizes low-rank retrieval performance but compromises top-rank retrieval performance.} 
Specifically, JointLH achieves the highest Recall@1000 but the lowest MRR@10 on the Dev. 
This confirms that its inherent mechanism distributes learning signals across the entire positive set rather than concentrating on the high-quality human label.

\subsection{Performance after Curriculum Learning} 
Curriculum learning (CL) can effectively combine LLM-annotated and human-annotated positives.
Concretely, we first train the retriever with each objective on the LLM-annotated MS~MARCO, and then fine-tune the resulting model on the human-annotated data. Results are shown in Table~\ref{tab:cl_msmarco}.
We can observe that 
(1) Most multi-positive loss objectives yield better performance than SingleLH, further indicating the necessity of high-quality multi-positive annotation. 
(2) The performance gaps among objectives become smaller after CL, suggesting that even a short refinement stage on high-quality human labels can bring measurable gains and make objective-specific differences smaller.

\begin{table}[t]
\centering
\caption{Retrieval performance on MS~MARCO after curriculum learning fine-tuning. R@1k and N@10 are defined in Table~\ref{tab:nonuniform_results}. \textbf{Bold} and \underline{underline} indicate the best and second-best results. ``+'' and ``-'' indicate statistically significant improvements and drops compared to SingleLH, while $\clubsuit$ and $\diamondsuit$ indicate statistically significant differences compared to Rand1LH (two-sided paired $t$-test, $p < 0.05$).}
\label{tab:cl_msmarco}
\resizebox{\linewidth}{!}{%
\begin{tabular}{lcccccc}
\toprule
\multirow{2}{*}{\textbf{Objective}} &
\multicolumn{2}{c}{\textbf{Dev Set}} &
\textbf{DL19} &
\textbf{DL20} &
\multicolumn{2}{c}{\textbf{Hybrid Annotation}} \\
\cmidrule(lr){2-3} \cmidrule(lr){4-4} \cmidrule(lr){5-5} \cmidrule(lr){6-7}
& \textbf{MRR@10} & \textbf{R@1k} & \textbf{N@10} & \textbf{N@10} & \textbf{MRR@10} & \textbf{N@10} \\
\midrule
SingleLH   & 33.65 & 96.00 & 62.34 & 63.60 & 80.14 & 54.32 \\
Rand1LH    & \textbf{34.13}\sigp & 96.21 & \underline{63.25} & 63.42 & \underline{82.04} & \underline{55.21} \\
JointLH    & 33.83 & \textbf{96.48}\btrs & 62.83 & \underline{64.69} & \textbf{83.83}\sigp & \textbf{55.70}\sigp \\
SumMargLH  & 33.08\wtrs & 95.97\wtr & 61.20 & 63.90 & 81.00 & 54.09 \\
LSEPair    & \underline{33.97} & \underline{96.42}\sigp & \textbf{65.35}\sigp & \textbf{64.91} & 80.67 & \textbf{55.70} \\
\bottomrule
\end{tabular}%
}
\end{table}

\section{Further Analysis}
\label{section:Further_Analysis}

\subsection{Impact of Positive-Negative Ratio}

Beyond the default setting (group size for dense retrieval training $G=8$ and max positive count in the training group $M=4$), we analyze how multi-positive objectives respond to the positive-negative ratio within each training group. 
On LLM annotated MS~MARCO, we vary (i) the group size $G \in \{8,16\}$ and (ii) the maximum number of positives per query used in training, denoted by $M$ (i.e., an upper bound on $|D^+|$). We evaluate $M\in\{2,4\}$ for $G{=}8$ and $M\in\{2,4,8\}$ for $G{=}16$. 
All other training hyperparameters are kept identical to our main MS~MARCO experiment (\S\ref{sec:res_uniform}).

\begin{figure}[t]
    \centering
    \makebox[\linewidth][c]{\includegraphics[width=1.0\linewidth]{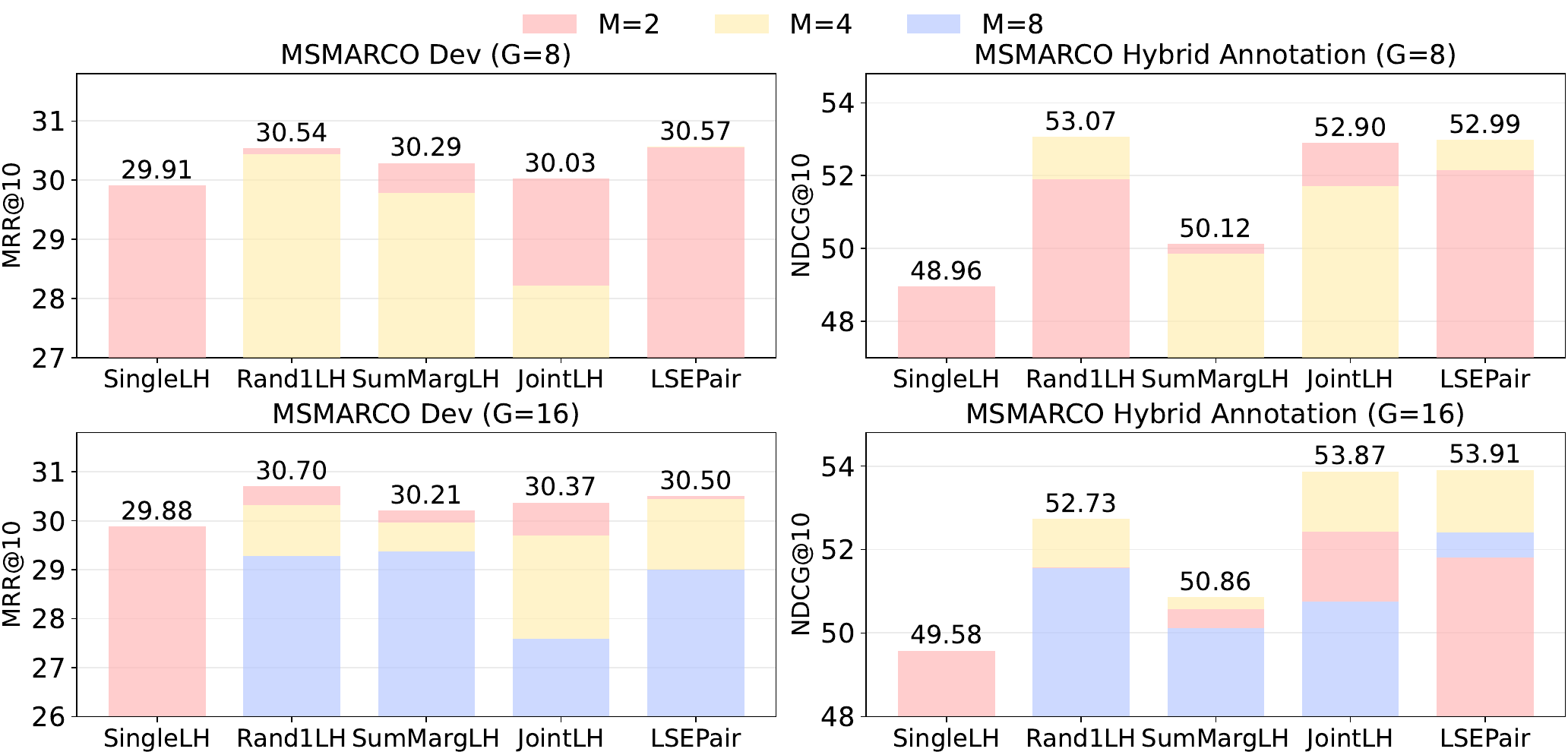}}
    \caption{Positive-negative ratio ablation on MS~MARCO by varying group size $G$ and the maximum positive count $M$.}
    \label{fig:ratio_ablation}
\end{figure}

Figure~\ref{fig:ratio_ablation} reports MRR@10 on Dev and NDCG@10 on the Hybrid Annotation set. 

We can observe that: 
(1) SumMargLH and LSEPair perform more stably compared to other objectives, whereas JointLH shows the most pronounced performance fluctuations as $M$ increases. 
The reason may be that SumMargLH optimizes the positive with the highest score during training, shown in Equation \eqref{eq:summar-grad};
adding weaker positives may not influence the score of the top positive. In contrast, JointLH is highly sensitive to the positive count $M$. For instance, at $G=16$, as $M$ increases from 2 to 8, JointLH experiences a sharp decline in MRR@10 from 30.37\% to 27.59\%. This confirms that JointLH's mechanism of distributing probability mass across an expanding set of positives inherently dilutes the model's top-rank sharpness, suggesting that JointLH may be less suitable for retrieval settings that require strong discrimination at the top ranks.
Since increasing $M$ also introduces additional positives that may be lower-quality, Figure~\ref{fig:ratio_ablation} alone does not fully disentangle the effect of the positive-negative ratio from the effect of label noise; we provide an additional controlled analysis in Appendix~\ref{sec:appendix_ratio_control}.
(2) Excessively large $M$ consistently degrades top-tier ranking effectiveness across all objectives. While increasing $M$ from 2 to 4 can benefit retrieval breadth (e.g., JointLH's Hybrid NDCG@10 rising from 52.43\% to 53.87\% at $G=16$), a further increase to $M=8$ leads to a universal performance drop. Notably, at $G=16$ and $M=8$, even the LSEPair and Rand1LH see their MRR@10 fall to 29.00\% and 29.28\% respectively, underperforming the SingleLH baseline (29.88\%). For LSEPair in particular, this degradation is consistent with the risk suggested by its gradient property: because low-scoring positives receive stronger updates, introducing false or marginal positives can misdirect the optimization signal. 
This confirms that when the positive-to-negative ratio becomes too high, the core supervision signal is diluted by an excess of lower-quality positives, preventing the model from forming the peaky ranking distribution necessary for optimal top-rank performance. 

\subsection{LSEPair's Variants for Dense Retrieval} 
After extensive experiments, we observe that LSEPair consistently delivers strong and robust performance across most settings. 
In this section, we explore the effectiveness of the Log-Sum-Exp Pairwise (LSEPair) loss and its variants for dense retrieval. 
The original LSEPair definition expands the aggregation to all pairs of positive and negative documents. 
We introduce several variants of the LSEPair objective by restricting the aggregation scope:
\begin{itemize}[leftmargin=*,itemsep=0pt,topsep=0pt,parsep=0pt]
    \item Max Positive Variant, which only considers the positive document with the highest score:
  $$\mathcal{L}_{\text{LSEPair\_maxP}} = \log \left( 1 + \sum_{d^- \in D^-} \exp(s(q, d^-) - s(q, d^+_{\mathrm{max}})) \right),$$
  where $d^+_{\mathrm{max}} = \arg\max_{d^+ \in D^+} s(q, d^+)$. 
  \item Max Negative Variant, which only considers the negative document with the highest score:  $$\mathcal{L}_{\text{LSEPair\_maxN}} = \log \left( 1 + \sum_{d^+ \in D^+} \exp(s(q, d^-_{\mathrm{max}}) - s(q, d^+)) \right),$$ where $d^-_{\mathrm{max}} = \arg\max_{d^- \in D^-} s(q, d^-)$.
  \item Min Positive Variant, which only considers the positive document with the lowest score: 
 $$\mathcal{L}_{\text{LSEPair\_minP}} = \log \left( 1 + \sum_{d^- \in D^-} \exp(s(q, d^-) - s(q, d^+_{\mathrm{min}})) \right),$$ where $d^+_{\mathrm{min}} = \arg\min_{d^+ \in D^+} s(q, d^+)$.

  \item Min Positive and Max Negative Variant, which only considers the positive document with the lowest score and the negative document with the highest score:   
 $$\mathcal{L}_{\text{LSEPair\_minP\_maxN}} = \log \left( 1 + \exp(s(q, d^-_{\mathrm{max}}) - s(q, d^+_{\mathrm{min}})) \right).$$
\end{itemize}
Retrieval performance is evaluated on NQ, MS~MARCO Dev, and BEIR, as shown in Table~\ref{tab:lsepair_variants}. 

\begin{table}[t]
\centering 

\caption{Performance of different LSEPair variants and SumMargLH training objectives.
\textbf{Bold} denotes the best result. ``+'' and ``-'' indicate significant differences compared to the reference strategy $\mathcal{L}_{\text{LSEPair\_maxP}}$ and $\mathcal{L}_{\text{LSEPair}}$, respectively.}
\label{tab:lsepair_variants}
\resizebox{\linewidth}{!}{%
\begin{tabular}{lccccc}
\toprule
\multirow{2}{*}{\textbf{Objectives}} & \multicolumn{2}{c}{\textbf{NQ}} & \multicolumn{2}{c}{\textbf{MS~MARCO Dev}} & \textbf{BEIR} \\
\cmidrule(lr){2-3} \cmidrule(lr){4-5} \cmidrule(lr){6-6}
& \textbf{Acc@20} & \textbf{Acc@100} & \textbf{MRR@10} & \textbf{R@1000} & \textbf{NDCG@10} \\
\midrule
$\mathcal{L}_{\text{LSEPair}}$ & \textbf{77.01}\sigp  & 85.62\sigp & \textbf{30.57} & 94.82 & \textbf{39.49} \\
$\mathcal{L}_{\text{SumMargLH}}$ & 75.12\sigp & 84.57\sigp & 29.79\sign & 92.98\sign & 37.10 \\
\midrule
$\mathcal{L}_{\text{LSEPair\_maxP}}$ & 74.18\sign & 83.91\sign & 29.78\sign & 92.68\sign & 36.86 \\
$\mathcal{L}_{\text{LSEPair\_maxN}}$ & 76.76\sigp & 85.79\sigp & 30.45\sigp & 94.00\sigp & 37.88 \\
$\mathcal{L}_{\text{LSEPair\_minP}} $ & 76.79\sigp & \textbf{85.82}\sigp & 30.32 & \textbf{95.05}\sigp & 39.16 \\
$\mathcal{L}_{\text{LSEPair\_minP\_maxN}}$ & 76.48\sigp & 85.76\sigp & 29.63\sign & 93.34\sign & 35.50 \\

\bottomrule
\end{tabular}
}
\end{table}
We observe that $\mathcal{L}_{\text{LSEPair\_maxP}}$ consistently underperforms the original LSEPair, suggesting that concentrating updates on the highest-scoring positive can hurt retrieval quality. 
For completeness, we also report the results of $\mathcal{L}_{\text{SumMargLH}}$ in Table~\ref{tab:lsepair_variants} (its main performance is reported in Table~\ref{tab:indomain_results}). 
Consistent with our earlier findings, $\mathcal{L}_{\text{SumMargLH}}$ also underperforms, further supporting the hypothesis that objectives which emphasize only the strongest positives may neglect lower-scoring yet valid positives.
Among the variants, $\mathcal{L}_{\text{LSEPair\_minP}}$ stays closest to the original LSEPair across datasets, achieving the best NQ Acc@100 and MS~MARCO Dev Recall@1000 while remaining competitive on BEIR. 
This indicates that lower-scoring positives provide useful training signals, while the full LSEPair aggregation remains the most reliable choice for balancing top-rank effectiveness and out-of-domain robustness.

\subsection{Annotation Budget Allocation}
As LLM-based relevance annotation becomes increasingly feasible, it is important to understand how to allocate a fixed annotation budget effectively. We therefore compare two annotation strategies: labeling more queries with fewer positives per query versus labeling fewer queries with more positives per query. In our analysis, we use the total number of positive labels as a practical proxy to simulate a fixed annotation budget. Although real annotation workflows may assign different costs to labeling a new query and identifying an additional positive passage for an already labeled query, this approximation remains informative for our controlled comparison.

Based on the LLM-annotated MS~MARCO pool, we compare four annotation settings: (i) all queries with a single positive each, (ii) a random 50\% subset of queries with the top two positives, (iii) a random one-third subset of queries with the top three positives, and (iv) a random 25\% subset of queries with four positives per query. 
All models are trained with LSEPair. 
We train for $3$, $4$, $5$, and $5$ epochs for the four settings, respectively. 
Table~\ref{tab:budget_query_vs_pos} reports MS~MARCO Dev performance. 
Under a fixed positive-label budget, allocating labels to \emph{more queries} (smaller $\mathrm{m}$) is more cost-effective for head ranking quality: Dev MRR@10 decreases as $\mathrm{m}$ increases, while Recall@1000 changes only marginally. However, using the top 4 positives and the entire query set achieves much better overall performance. 
This suggests that when annotation is expensive (e.g., human labeling), it is preferable to expand query coverage rather than annotate many positives per query; in contrast, for low-cost LLM annotation, it would be better to scale both query coverage and per-query depth of the annotation pool.
\begin{table}[t]
\centering
\caption{Retrieval performance under a fixed positive-label budget. \textbf{Bold} and \underline{underline} denote the best and second-best results, respectively. ``QCR'' means the query count ratio.}
\label{tab:budget_query_vs_pos}
\begin{tabular}{crcc}
\toprule
\multirow{2}{*}{\textbf{Positive Count}} & \multirow{2}{*}{\textbf{QCR}} &  \multicolumn{2}{c}{\textbf{MS MARCO Dev}} \\
\cmidrule(lr){3-4}
& & \textbf{MRR@10} & \textbf{Recall@1000} \\
\midrule
4 & 100\%  & \textbf{30.57} & \textbf{94.82} \\
\midrule
1 & 100\%  & \underline{29.83} & 93.60 \\
2 & 50\%   & 29.50 & \underline{93.80} \\
3 & 33.3\% & 27.81 & 93.65 \\
4 & 25\%   & 26.57 & 93.67 \\
\bottomrule
\end{tabular}
\end{table}
\vfill\eject
\section{Conclusion and Future Work}
\label{sec:Conclusion}
In this work, we systematically investigate how to effectively leverage multi-positive supervision for dense retrieval, a scenario increasingly enabled by scalable LLM-based annotations. 
By unifying a range of listwise and pairwise objectives within a common contrastive learning framework, we characterize their mathematical relationships, distinct gradient behaviors, and inductive biases. 
Empirical evaluations on multiple datasets with both LLM-annotated and mixed human-LLM labels confirm our theoretical findings. LSEPair emerges as a consistently strong objective, while Rand1LH proves to be a reliable and simple baseline. 
In contrast, SumMargLH and JointLH are more sensitive to the distribution and quality of positives. 
In summary, multi-positive supervision is not merely an increase in label quantity, but a qualitatively different training signal that requires thoughtful objective and data construction. Our work provides guidance for retriever training under multi-positive supervision and lays the foundation for future research on learning objectives in scenarios with heterogeneous labels. 

While our work investigates multi-positive supervision in dense retrieval, an important direction for future work is to extend this analysis to broader ranking tasks. Ranking has long used objectives that can naturally handle multiple positives, especially pairwise losses over positive-negative document pairs. However, dense retrieval differs from many ranking scenarios. Studying how the same objectives behave when the candidate set is fixed or densely judged would clarify which findings are specific to dense retrieval and which generalize to ranking more broadly.

\begin{acks}
This work was funded by the National Natural Science Foundation of China (NSFC) under Grants No. 62302486 and No. 62441229, the Innovation Project of ICT CAS under Grants No. E361140, and the Strategic Priority Research Program of the CAS under Grants No. XDB0680102.
\end{acks}


\clearpage

\bibliographystyle{ACM-Reference-Format}
\balance
\bibliography{sample-base}

\appendix

\section{Detailed Gradient Derivations}

In this section, we provide the detailed derivation steps for the gradients of the proposed multi-positive objectives with respect to the score of a positive document $s(q, d^+_i)$.
For brevity, we denote $s_i = s(q, d^+_i)$, $s_j = s(q, d^+_j)$ for positive documents, and $s_k = s(q, d^-_k)$ for negative documents.
Let $Z = \sum_{d \in \mathcal{D}} \exp(s(q, d))$ denote the normalization term (partition function) over the entire candidate set $\mathcal{D} = D^+ \cup D^-$.

\subsection{Derivation for JointLH}
The JointLH objective is defined as the average negative log-likelihood over the positive set $D^+$:
\begin{equation}
\mathcal{L}_{\text{JointLH}} = - \frac{1}{|D^+|} \sum_{d^+_j \in D^+} \log \frac{\exp(s_j)}{Z}.
\end{equation}
Expanding the logarithmic term:
\begin{equation}
\mathcal{L}_{\text{JointLH}} = - \frac{1}{|D^+|} \sum_{d^+_j \in D^+} s_j + \log Z.
\end{equation}
The gradient with respect to a specific positive score $s_i$ is:
\begin{equation}
\begin{aligned}
\frac{\partial \mathcal{L}_{\text{JointLH}}}{\partial s_i} &= - \frac{1}{|D^+|} \frac{\partial}{\partial s_i} \left( \sum_{d^+_j \in D^+} s_j \right) + \frac{\partial \log Z}{\partial s_i} \\
&= - \frac{1}{|D^+|} \cdot 1 + \frac{1}{Z} \frac{\partial Z}{\partial s_i} \\
&= - \frac{1}{|D^+|} + \frac{\exp(s_i)}{Z} \\
&= P(d^+_i|q, \mathcal{D}) - \frac{1}{|D^+|}.
\end{aligned}
\end{equation}

\subsection{Derivation for SumMargLH}
The SumMargLH objective maximizes the marginal probability of the positive set. Let $Z^+ = \sum_{d^+_j \in D^+} \exp(s_j)$ be the sum of positive exponentiated scores. The loss is:
\begin{equation}
\mathcal{L}_{\text{SumMargLH}} = - \log \frac{Z^+}{Z} = - \log Z^+ + \log Z.
\end{equation}
The gradient with respect to $s_i$ is derived as:
\begin{equation}
\begin{aligned}
\frac{\partial \mathcal{L}_{\text{SumMargLH}}}{\partial s_i} &= - \frac{\partial \log Z^+}{\partial s_i} + \frac{\partial \log Z}{\partial s_i} \\
&= - \frac{1}{Z^+} \frac{\partial Z^+}{\partial s_i} + \frac{1}{Z} \frac{\partial Z}{\partial s_i} \\
&= - \frac{\exp(s_i)}{Z^+} + \frac{\exp(s_i)}{Z} \\
&= - \frac{\exp(s_i) \cdot Z}{Z^+ \cdot Z} + \frac{\exp(s_i)}{Z} \\
&= - \frac{\exp(s_i)}{Z} \cdot ( \frac{Z}{Z^+} - 1 ).
\end{aligned}
\end{equation}
Rearranging terms highlights the gradient scaling:
\begin{equation}
\frac{\partial \mathcal{L}_{\text{SumMargLH}}}{\partial s_i} = 
- \frac{\exp(s_i)}{\sum_{d \in \mathcal{D}} \exp(s(q, d))} \cdot
(\frac{\sum_{d \in \mathcal{D}} \exp(s(q, d))}{\sum_{d^+_j \in D^+}\exp s(q, d^+)} - 1)
\end{equation}

\subsection{Derivation for LSEPair}
The LSEPair objective aggregates all pairwise constraints:
\begin{equation}
\mathcal{L}_{\text{LSEPair}} = \log \left( 1 + \sum_{d^+_j \in D^+} \sum_{d^-_k \in D^-} \exp(s_k - s_j) \right).
\end{equation}
Let $\Omega = 1 + \sum_{d^+_j \in D^+} \sum_{d^-_k \in D^-} \exp(s_k - s_j)$ be the argument of the logarithm.
The gradient with respect to $s_i$ is:
\begin{equation}
\begin{aligned}
\frac{\partial \mathcal{L}_{\text{LSEPair}}}{\partial s_i} &= \frac{1}{\Omega} \cdot \frac{\partial \Omega}{\partial s_i} \\
&= \frac{1}{\Omega} \cdot \frac{\partial}{\partial s_i} \left( \sum_{d^-_k \in D^-} \exp(s_k - s_i) \right) \\
&= -\frac{1}{\Omega} \cdot \sum_{d_k^-\in D^-}\exp (s_k - s_i) \\
&= -\frac{\exp(-s_i) \cdot \sum_{d_k^-\in D^-}\exp (s_k)}{\Omega} .
\end{aligned}
\end{equation}
This confirms that the gradient magnitude is proportional to $\exp(-s_i)$, prioritizing lower-scoring positives.

\section{Controlled Analysis of Positive-Negative Ratio}
\label{sec:appendix_ratio_control}
In Section~\ref{section:Further_Analysis}, increasing the maximum number of positives $M$ changes two factors: it raises the positive-negative ratio in the training group and may introduce lower-quality positives. To better separate these effects, we conduct an additional controlled experiment on MS~MARCO with group size fixed to $G=16$. Besides the first-$M$ setting used in the main analysis, we also randomly sample $M$ positives from the available positive pool for each query. This random-$M$ construction weakens the dependence on the utility-based positive ordering, making it easier to examine whether increasing the positive count itself contributes to the performance drop.

\begin{table}[t]
\centering
\caption{Controlled analysis of positive selection under $G=16$ on MS~MARCO. ``First $M$'' uses the top-$M$ positives from the annotation pipeline, while ``Random $M$'' samples $M$ positives from the available positive pool.}
\label{tab:appendix_ratio_control}
\resizebox{\linewidth}{!}{%
\begin{tabular}{lccccc}
\toprule
\multirow{2}{*}{\textbf{Objective}} & \multirow{2}{*}{\textbf{$M$}} & \multicolumn{2}{c}{\textbf{First $M$}} & \multicolumn{2}{c}{\textbf{Random $M$}} \\
\cmidrule(lr){3-4} \cmidrule(lr){5-6}
& & \textbf{Dev MRR@10} & \textbf{Hybrid NDCG@10} & \textbf{Dev MRR@10} & \textbf{Hybrid NDCG@10} \\
\midrule
JointLH & 4 & 29.70 & 53.87 & 27.65 & 51.58 \\
JointLH & 8 & 27.59 & 50.75 & 27.01 & 50.80 \\
SumMargLH & 4 & 29.97 & 50.86 & 28.97 & 49.50 \\
SumMargLH & 8 & 29.37 & 50.11 & 28.84 & 49.72 \\
LSEPair & 4 & 30.45 & 53.91 & 28.33 & 51.65 \\
LSEPair & 8 & 29.00 & 52.41 & 28.12 & 51.51 \\
\bottomrule
\end{tabular}%
}
\end{table}

Table~\ref{tab:appendix_ratio_control} shows that JointLH still degrades when $M$ increases from 4 to 8 under random positive sampling. This suggests that the positive-negative ratio change itself contributes to the performance drop, rather than the drop being solely caused by adding lower-ranked positives in the first-$M$ setting. At the same time, the decline is larger in the first-$M$ setting, and SumMargLH and LSEPair show much smaller changes under random sampling. These results indicate that label quality and noise also affect the observed degradation, especially when the additional positives include marginally relevant or false-positive passages.

\section{Dynamic Positive Reweighting Loss}
\label{sec:appendix_dynamic_weighting}

ReContriever~\cite{lei-etal-2023-unsupervised} proposes relevance-aware contrastive pre-training for unsupervised dense retrieval, where the model estimates the relevance of pseudo-positive pairs and adaptively reweights the contrastive objective. This differs from our supervised multi-positive setting, where each query is associated with explicitly labeled positive passages. Nevertheless, since dynamic weighting provides another way to aggregate multiple positive signals, we adapt this idea as an additional loss by assigning positives query-dependent weights according to their estimated relevance scores.

\begin{table}[t]
\centering
\caption{Comparison with a dynamic positive reweighting baseline on MS~MARCO and BEIR. The ReContriever-style baseline adapts relevance-aware positive weighting to our supervised multi-positive setting.}
\label{tab:appendix_dynamic_weighting}
\resizebox{\linewidth}{!}{%
\begin{tabular}{lccccc}
\toprule
\multirow{2}{*}{\textbf{Objective}} & \multicolumn{2}{c}{\textbf{MS~MARCO Dev}} & \textbf{DL19} & \textbf{DL20} & \textbf{BEIR} \\
\cmidrule(lr){2-3} \cmidrule(lr){4-4} \cmidrule(lr){5-5} \cmidrule(lr){6-6}
& \textbf{MRR@10} & \textbf{Recall@1000} & \textbf{NDCG@10} & \textbf{NDCG@10} & \textbf{NDCG@10} \\
\midrule
JointLH & 28.22 & 95.10 & 59.81 & 63.20 & 37.67 \\
LSEPair & 30.57 & 94.82 & 65.33 & 62.80 & 39.49 \\
ReContriever & 28.30 & 95.14 & 60.26 & 62.87 & 37.33 \\
\bottomrule
\end{tabular}%
}
\end{table}

As shown in Table~\ref{tab:appendix_dynamic_weighting}, the ReContriever-style dynamic weighting baseline is competitive with JointLH, slightly improving MS~MARCO Dev MRR@10 from 28.22 to 28.30 and Recall@1000 from 95.10 to 95.14. However, it remains substantially behind LSEPair on top-rank effectiveness and out-of-domain performance, especially on MS~MARCO Dev MRR@10 (28.30 vs. 30.57), DL19 NDCG@10 (60.26 vs. 65.33), and BEIR average NDCG@10 (37.33 vs. 39.49). These results suggest that dynamic relevance-aware weighting is a useful baseline for multi-positive aggregation, but in our supervised dense retrieval setting it behaves closer to JointLH than to LSEPair and does not replace the benefit of explicitly optimizing positive-negative pairwise separation.

\end{document}